# An Empirical Biomarker-based Calculator for Autosomal Recessive Polycystic Kidney Disease

## The Nieto-Narayan Formula

Jake A. Nieto, Michael A. Yamin, Itzhak D. Goldberg and Prakash Narayan*

*Corresponding Author
Address: Angion Biomedica Corp., 51 Charles Lindbergh Blvd, Uniondale, 11553, New York, USA
Ph: 516 326 1200
Fax: 516 222 1359
Email: pnarayan@angion.com

**Running Title**: Biomarker-based Calculator for ARPKD

## Abstract

Autosomal polycystic kidney disease (ARPKD) is associated with progressive enlargement of the kidneys fuelled by the formation and expansion of fluid-filled cysts. The disease is congenital and children that do not succumb to it during the neonatal period will, by age 10 years, more often than not, require nephrectomy+renal replacement therapy for management of both pain and renal insufficiency. Since increasing cystic index (CI; percent of kidney occupied by cysts) drives both renal expansion and organ dysfunction, management of these

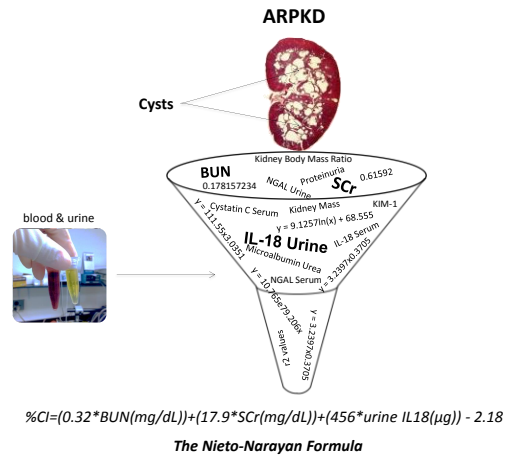

%CI=(0.32*BUN(mg/dL))+(17.9*SCr(mg/dL))+(456*urine IL18(μg)) - 2.18
*The Nieto-Narayan Formula*

patients, including decisions such as elective nephrectomy and prioritization on the transplant waitlist, could clearly benefit from serial determination of CI. So also, clinical trials in ARPKD evaluating the efficacy of novel drug candidates could benefit from serial determination of CI. Although ultrasound is currently the imaging modality of choice for diagnosis of ARPKD, its utilization for assessing disease progression is highly limited. Magnetic resonance imaging or computed tomography, although more reliable for determination of CI, are expensive, time-consuming and somewhat impractical in the pediatric population. Using a well-established mammalian model of ARPKD, we undertook a big data-like analysis of minimally- or non-invasive serum and urine biomarkers of renal injury/dysfunction to derive a family of equations for estimating CI. We then applied a signal averaging protocol to distill these equations to a single empirical formula for calculation of CI. Such a formula will eventually find use in identifying and monitoring patients at high risk for progressing to end-stage renal disease and aid in the conduct of clinical trials.

## Introduction

Autosomal recessive polycystic kidney disease (ARPKD) is a genetic disorder caused by a mutation in the polycystic kidney and hepatic disease 1 (*PKHD1*) gene and affects ~ 1 in 20,000 children [1-4]. Fluid-filled cyst formation and expansion replaces normal tissue and progressively increases kidney size [5,6]. Nevertheless, owing to the remarkable degree to which intact nephrons can compensate for loss of functioning parenchyma, glomerular function rate measurements fail to disclose the increasing cystic index (CI; percent of kidney occupied by cysts) until late in the disease [6]. In ARPKD, dialysis+nephrectomy and/or kidney transplantation are inevitable and are driven, both, by the need for pain management via renal volume reduction and restitution of renal function.

Aggressive use of available therapies (e.g. blood pressure lowering drugs) has improved survival in ARPKD. While 20-30% of infants with the disease still die hours or days after birth due to breathing difficulties, of those that survive infancy, ~ 82% survive to age 10 and ~ 73% live past the age of 15 years [3,4]. Management of this patient population including decisions such as elective nephrectomy and prioritization on the transplant waitlist could clearly benefit from serial determination of CI, the key driver of increased kidney volume and (eventual) renal failure. So also, clinical trials in ARPKD evaluating efficacy of novel drug candidates could clearly benefit from serial measurements of CI [7]. Although ultrasound is currently the imaging modality of choice for a diagnosis of ARPKD, its utilization for assessing disease progression is limited [8]. Due to the irregular shape and enormous number of cysts, only computed tomography (CT) or magnetic resonance imaging (MRI) can accurately measure cyst content [8]. Unfortunately such contrast-enhanced scanning modalities are expensive, time-consuming and somewhat impractical in the pediatric population.

In a well-established mammalian model of ARPKD [9,10], we have previously [9] determined CI while evaluating serum and urine levels of several renal injury/dysfunction biomarkers. We queried this database to identify and quantify a relationship, if any, between CI and this family of biomarkers.

## Methods

**Source Data:** No animals were used in the analysis of these data. Rather, all data analyzed in this study were sourced from a previously published study [9] characterizing a rodent model of ARPKD. In that study [9], male PCK (PCK/CrljCrl-Pkhd1pck/Crl) rats carrying the *PKHD1* mutation for ARPKD, had 1 kidney removed at ~10.5 weeks of age to accelerate renal dysfunction. Animals were sacrificed at age 13.5 weeks. Immediately prior to sacrifice, 24 hr urine was collected in metabolic cages and animals weighed. Blood was drawn at sacrifice and the kidney retrieved for analysis. Levels of a panel of kidney-relevant biomarkers [see Table 1] were analyzed using either enzyme-linked immunoabsorbent assay (*ELISA*), a core services laboratory (Northwell Health, NY) or biochemical assays [9]. Concentration of urine biomarkers were multiplied by 24 hr urine volume. CI (% cyst space/renal parenchyma) was measured from hematoxylin and eosin (H&E)-stained renal sections (two separate sections per kidney) by two independent observers and the values averaged so as to get a representative CI value for the kidney.

**Correlation Coefficient Criteria**: Microsoft Excel 2010 curve fitting software was used to generate plots of CI against renal mass, renal:body mass ratio and biomarker levels. For a given CI, if a corresponding biomarker level was missing, that pair was eliminated from the analysis. Final n values for each correlation are reported in Table 1. Correlation coefficient r, i.e. the simultaneous fluctuation occurring between two variables, was calculated from $r^2$ values generated off the trend line fitted using exponential, linear, logarithmic or polynomial regression. To determine whether r was significant, the r value and the sample size n were entered into an online calculator [11]. A p value <0.05 was deemed significant. An r $\geq$ 0.7 with a p<0.05 was used as the lower limit for CI vs. biomarker correlation.

## Results

Table 1 reports the variables, including biomarkers, correlated with CI.

| | Variable/Biomarker | Datapoints (n) |
|---|---|---|
| %CI | kidney mass (g) | 27 |
| | kidney/body mass ratio | 27 |
| | NGAL, serum (µg/mL) and urine (µg) | 24 and 25 |
| | KIM-1, urine (µg) | 12 |
| | Cystatin C, serum (µg/mL) and urine (µg) | 25 and 24 |
| | IL-18, serum (µg/mL) and urine (µg) | 27 and 23 |
| | SCr (mg/dL) | 27 |
| | BUN (mg/dL) | 27 |
| | proteinuria (mg) | 24 |
| | microalbuminuria (µg) | 24 |

**Table 1: CI and Renal Variables/Biomarkers.** CI was correlated with kidney mass, kidney/body mass ratio and serum and/or urine-based renal biomarkers including neutrophil gelatinase-associated lipocalin (NGAL), Kidney Injury Molecule-1 (KIM-1), Cystatin C, interleukin (IL)-18, serum creatinine (SCr), blood urea nitrogen (BUN), proteinuria and microalbuminuria. The n represents the number of datapoints available for a given biomarker and CI set.

Individual values for CI, kidney mass, kidney to body mass ratio and the biomarkers evaluated are listed in Table 2. We queried these source data to draw out and quantify relationships, if any, between CI and biomarkers.

| %CI | kidney mass (g) | kidney/body mass | NGAL serum (µg/mL) | NGAL urine (µg) | KIM-1 urine (µg/mL) | Cystatin C serum (µg/mL) | Cystatin C urine (µg) | IL-18 serum (µg/mL) | IL-18 urine (µg) | SCr (mg/dL) | BUN (mg/dL) | proteinuria (mg) | microalbuminuria (µ |
|---|---|---|---|---|---|---|---|---|---|---|---|---|---|
| 27.5792 | 6.7300 | 1.7526 | NA | 960.00 | 0.00268579 | 0.5727088 | 14.999287 | 0.000034 | 0.005328081 | 0.61 | 0.61 | 427.73 | 72.02727 |
| 30.3611 | 6.7500 | 1.6504 | 3.99 | 1088.00 | 0.00284102 | 1.129012 | 3.438369 | 0.000027 | 0.011396097 | 0.62 | 0.62 | 489.05 | 61.366056 |
| 30.9221 | 7.6000 | 1.8225 | 1.54 | 120.18 | 0.00285067 | 1.428816 | 14.513676 | 0.000032 | 0.009516466 | 0.78 | 0.78 | 434.90 | 79.9212015 |
| 34.4874 | 7.2500 | 1.6477 | 0.86 | 73.92 | 0.00331086 | 1.116876 | 7.827184 | 0.000078 | 0.02001416 | 0.5 | 0.5 | 527.10 | 91.84422 |
| 20.9282 | 5.4000 | 1.2135 | 0.71 | 168.53 | 0.00253043 | 1.3691936 | 9.64404 | 0.000026 | 0.005473673 | 0.48 | 0.48 | 563.21 | 104.99436 |
| 26.2359 | 7.1100 | 1.6969 | 1.18 | 636.69 | 0.00282213 | 0.792976 | 13.8921356 | 0.000036 | 0.006671703 | 0.49 | 0.49 | 409.87 | 55.147743 |
| 23.7730 | 5.5400 | 1.3781 | 1.28 | 629.03 | 0.00486893 | 1.8374668 | 14.6956032 | 0.000085 | 0.013753793 | 0.53 | 0.53 | 782.75 | 125.102412 |
| 26.0932 | 5.7300 | 1.2140 | 1.05 | 63.54 | 0.00283051 | 0.982504 | 21.323152 | 0.000100 | 0.009541357 | 0.54 | 0.54 | 618.30 | 134.44605 |
| 22.0227 | 4.8000 | 1.1241 | 1.05 | 416.73 | 0.0048303 | 1.2236956 | 23.996291 | 0.000056 | 0.008970211 | 0.49 | 0.49 | 562.65 | 113.025105 |
| 19.8303 | 5.0600 | 1.2650 | 0.82 | 206.11 | 0.00343274 | 1.314648 | 0.408704 | 0.000016 | 0.007020066 | 0.53 | 0.53 | 389.92 | 70.39584 |
| 19.6681 | 6.1600 | 1.3391 | 0.97 | 70.90 | 0.00476574 | 1.002896 | 16.7835192 | 0.000066 | 0.007412596 | 0.56 | 0.56 | 799.80 | 173.22354 |
| 15.4724 | 6.0600 | 1.4975 | 3.06 | 1216.00 | 0.00494137 | 1.8639308 | 19.4514438 | 0.000037 | 0.005764512 | 0.58 | 0.58 | 752.99 | 121.265334 |
| 21.0398 | 8.4100 | 1.9835 | 3.38 | NA | NA | 1.4352892 | NA | 0.000049 | NA | 0.63 | 0.63 | NA | NA |
| 25.2283 | 7.0000 | 1.9391 | NA | NA | NA | 2.4615308 | NA | 0.000134 | NA | 0.77 | 0.77 | NA | NA |
| 23.7605 | 6.700 | 1.567 | NA | 384.00 | NA | 1.3336624 | 3.7887408 | 0.000065 | 0.005986509 | 0.58 | 0.58 | 143.81 | 23.824395 |
| 11.5877 | 3.6200 | 0.921 | 1.92 | 432.01 | NA | 1.1168876 | 0.2828436 | 0.000048 | NA | 0.38 | 0.38 | 188.85 | 35.212266 |
| 5.9299 | 3.2800 | 0.8367 | 0.80 | 39.76 | NA | 0.974564 | 0.00950032 | 0.000019 | 0.002081549 | 0.43 | 0.43 | 138.42 | 30.274965 |
| 0.4484 | 3.4900 | 0.8135 | 0.73 | 65.88 | NA | 0.670196 | 2.923987 | 0.000022 | NA | 0.33 | 0.33 | 307.97 | 67.96251 |
| 17.3568 | 5.0900 | 1.1211 | 1.08 | 66.85 | NA | 0.5895676 | 0.17662208 | 0.000026 | 0.002683384 | 0.47 | 0.47 | 208.96 | 50.052042 |
| 3.8756 | 4.000 | 0.9412 | 2.01 | 381.62 | NA | 0.6841508 | 3.8166282 | 0.000056 | 0.001771726 | 0.38 | 0.38 | 287.14 | 47.789838 |
| 7.5341 | 5.3000 | 1.2619 | 1.17 | 133.07 | NA | 1.4617588 | 0.10813803 | 0.000119 | 0.000187699 | 0.54 | 0.54 | 147.88 | 25.6407375 |
| 18.9592 | 4.9800 | 1.1116 | 1.78 | 35.91 | NA | 1.2026532 | 1.46491344 | 0.000049 | 0.00218640.4 | 0.52 | 0.52 | 260.17 | 35.50815 |
| 18.0817 | 5.7100 | 1.4134 | 1.70 | 355.69 | NA | 0.55031 | 4.188086 | 0.000074 | 0.002382867 | 0.55 | 0.55 | 238.80 | 31.93545 |
| 16.8749 | 7.7400 | 1.7671 | 1.25 | 184.60 | NA | 1.66142 | 7.4632298 | 0.000021 | 0.002955052 | 0.58 | 0.58 | 292.41 | 36.658869 |
| 13.7136 | 5.6100 | 1.249 | 1.21 | 40.07 | NA | NA | 0.254075 | 0.000062 | 0.0019058 | 0.56 | 0.56 | 215.03 | 41.81178 |
| 10.6465 | 4.5200 | 1.060 | 1.44 | 300.97 | NA | 0.88 | 8.1518544 | 0.000032 | 0.002226449 | 0.38 | 0.38 | 252.14 | 61.582885 |
| 12.0326 | 5.9400 | 1.2122 | 1.24 | NA | NA | NA | NA | 0.000025 | NA | 0.46 | 0.46 | NA | NA |

**Table 2: Source Data.** Data from a published study (9) were used as the source data for identifying and quantitating potential relationships between CI and renal biomarkers. NA=not available.

Since an increasing CI or ARPKD disease progression results in kidney enlargement and increased kidney mass [5], we first sought to confirm such a relationship between kidney mass and CI in this ARPKD dataset. As seen in Figures 1A and 1B, an excellent correlation is observed between CI and kidney mass. While the data could theoretically be fit by a number of regression strategies, each yielding an r ≥ 0.7 and a p<0.01 (Figure 1B), a linear relation between CI and kidney mass was clearly observed. The Pearson product-moment correlation coefficient, r, a measure of the strength and direction of the linear relationship, yielded a value of 0.73 with a p<0.01.

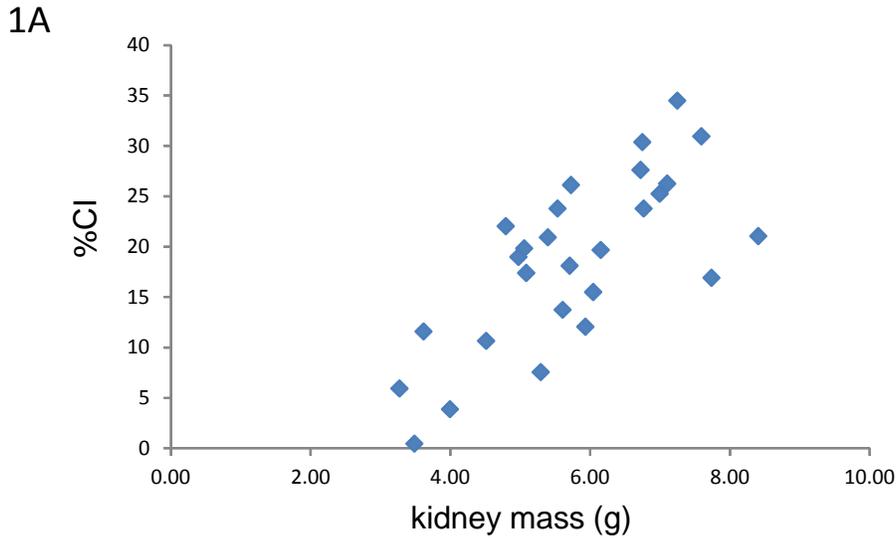

| Regression/Type | Equation | $r^2$ | r | p |
|---|---|---|---|---|
| Linear | y = 4.63x - 8 | 0.53 | 0.73 | <.01 |
| Logarithmic | y = 25.94ln(x) - 26.03 | 0.56 | 0.75 | <.01 |
| Polynomial | y = -0.45$x^3$ + 6.64$x^2$ - 25.85x + 34.02 | 0.61 | 0.78 | <.01 |
| Power | y = 0.21$x^{2.47}$ | 0.49 | 0.70 | <.01 |
| Exponetial | y = 1.31$e^{0.43x}$ | 0.43 | 0.65 | <.01 |

**Figure 1: CI and Kidney Mass.** CI tracks renal mass across a range of values (A).  A linear correlation is also observed between these 2 variables across the CI spectrum (B).

Next, CI was correlated with kidney to body mass ratio. Once again, excellent correlation was observed between these two variables (Figures 2A and 2B). Similar to its relation with renal mass, a linear relationship was observed between CI and kidney to body mass ratio with a Pearson product-moment correlation coefficient, r of 0.73 with a p<0.01.

.

2A

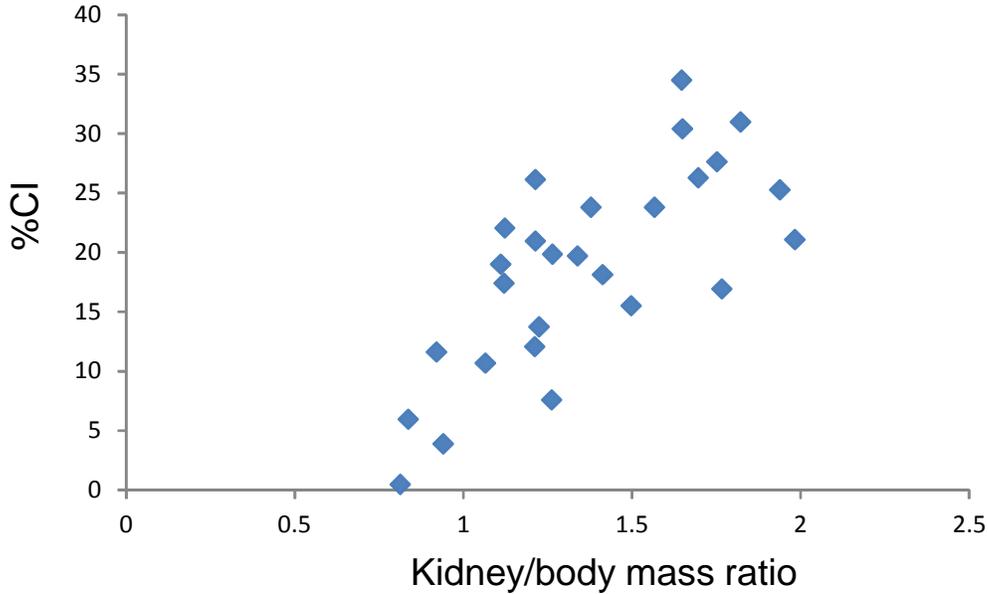

2B

| Regression/Type | Equation | $r^2$ | r | p |
|---|---|---|---|---|
| Linear | y = 18.52x - 6.54 | 0.53 | 0.73 | <.01 |
| Logarithmic | y = 25.72ln(x) + 11.5 | 0.58 | 0.76 | <.01 |
| Polynomial | y = -23.50x$^2$ + 84.13x - 49.8 | 0.62 | 0.79 | <.01 |
| Power | y = 7.73x$^{2.42}$ | 0.49 | 0.70 | <.01 |
| Exponetial | y = 1.58e$^{1.66x}$ | 0.41 | 0.64 | <.01 |

**Figure 2: CI and Kidney to Body Mass Ratio.** CI tracks kidney to body mass ratio across a range of values (A). A linear correlation is also observed between these 2 variables across the CI spectrum (B).

Serum and/or urine-based biomarkers of renal injury/dysfunction were then correlated with CI. The majority of biomarkers listed in Table 2 gave poor or no correlation with CI. An example of this is the lack of correlation between CI and serum Cystatin C as shown in Figures 3A and B.

3A

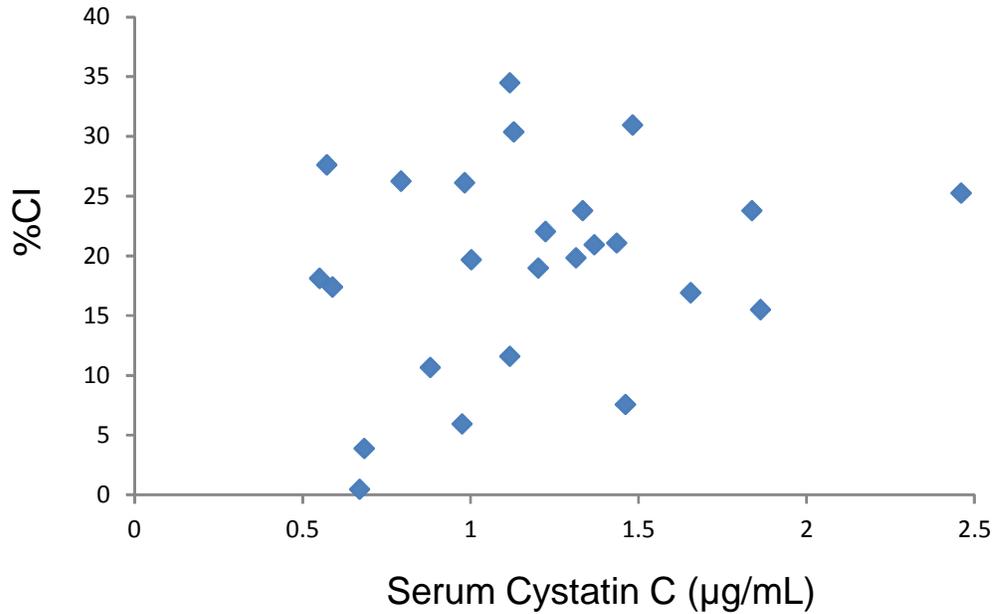

3B

| Regression/Type | Equation | $r^2$ | r | p |
|---|---|---|---|---|
| Linear | y = 4.54x + 13.76 | 0.06 | 0.24 | NS |
| Logarithmic | y = 5.53ln(x) + 18.6 | 0.06 | 0.25 | NS |
| Polynomial | y = -2.15x$^2$ + 10.38x + 10.3 | 0.06 | 0.25 | NS |
| Power | y = 14.27x$^{0.76}$ | 0.11 | 0.33 | NS |
| Exponetial | y = 7.47e$^{0.61x}$ | 0.10 | 0.31 | NS |

**Figure 3: CI and Serum Cystatin C.** There is no correlation between CI and serum Cystatin C in this model of ARPKD (A and B).

By contrast, as seen in Figures 4, 5 and 6, BUN, SCr and 24 hr urine IL-18 values, respectively, demonstrated excellent correlation with CI.

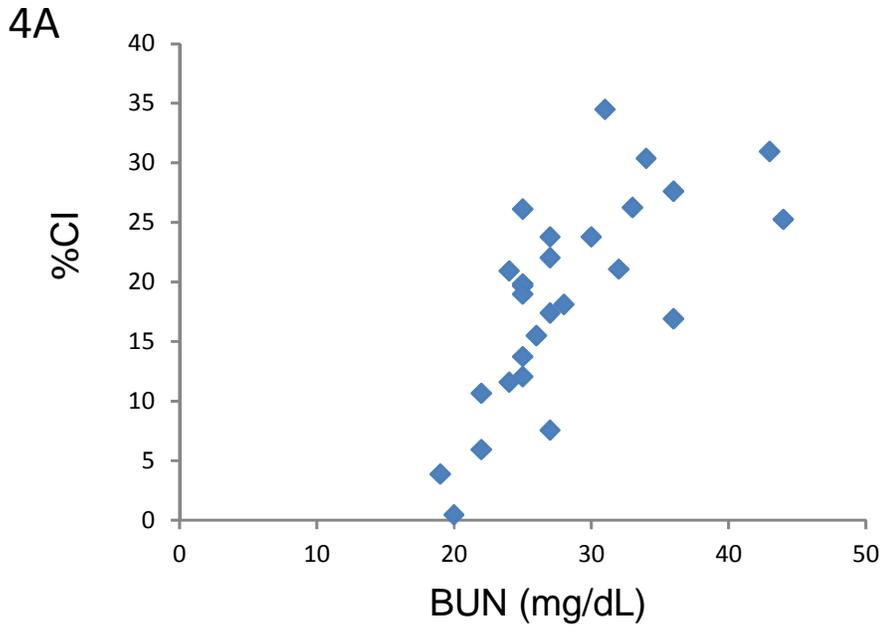

### 4A

### 4B

| Regression/Type | Equation | $r^2$ | r | p |
|---|---|---|---|---|
| Linear | y = 0.95x - 8.15 | 0.48 | 0.70 | <.01 |
| Logarithmic | y = 30.3ln(x) - 81.85 | 0.55 | 0.74 | <.01 |
| Polynomial | y = -0.07$x^2$ + 5.19x - 71.46 | 0.63 | 0.79 | <.01 |
| Power | y = 0.002$x^{2.67}$ | 0.40 | 0.64 | <.01 |
| Exponetial | y = 1.58$e^{0.08x}$ | 0.33 | 0.57 | <.01 |

**Figure 4: CI and BUN.** CI tracks BUN across a range of values (A). A linear correlation is also observed between these 2 variables across the CI spectrum (B).

5A

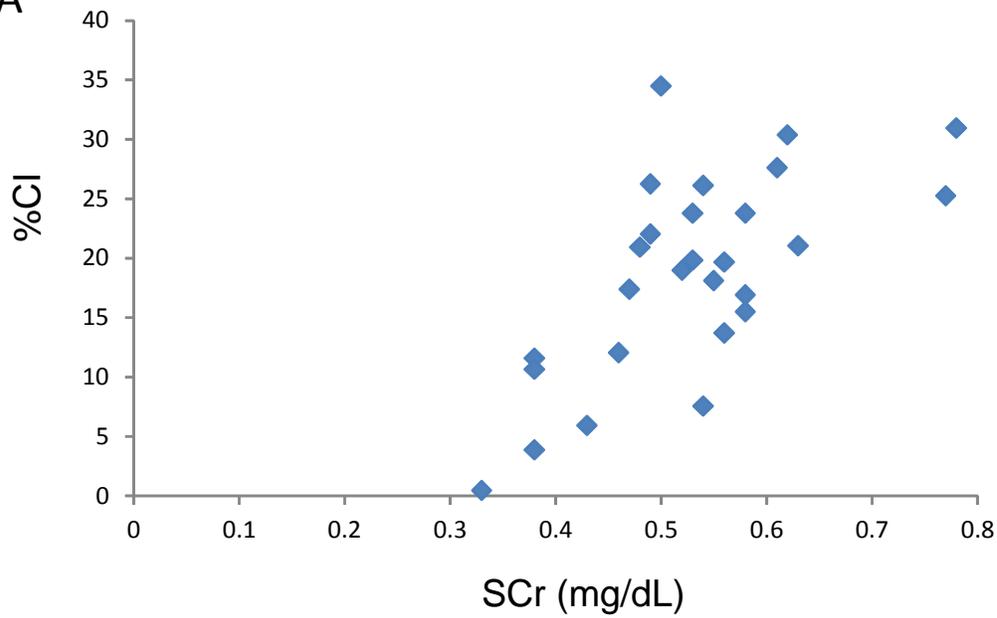

5B

| Regression/Type | Equation | $r^2$ | r | p |
|---|---|---|---|---|
| Linear | y = 53.63x - 9.66 | 0.44 | 0.66 | <.01 |
| Logarithmic | y = 29.45ln(x) + 38.03 | 0.48 | 0.69 | <.01 |
| Polynomial | y = -125.47$x^2$ + 192.12x - 46.48 | 0.50 | 0.71 | <.01 |
| Power | y = 111.55$x^{3.04}$ | 0.49 | 0.70 | <.01 |
| Exponetial | y = 0.95$e^{5.24x}$ | 0.40 | 0.63 | <.01 |

**Figure 5: CI and SCr.** CI tracks SCr across a range of values (A). A linear correlation is also observed between these 2 variables across the CI spectrum (B).

6A

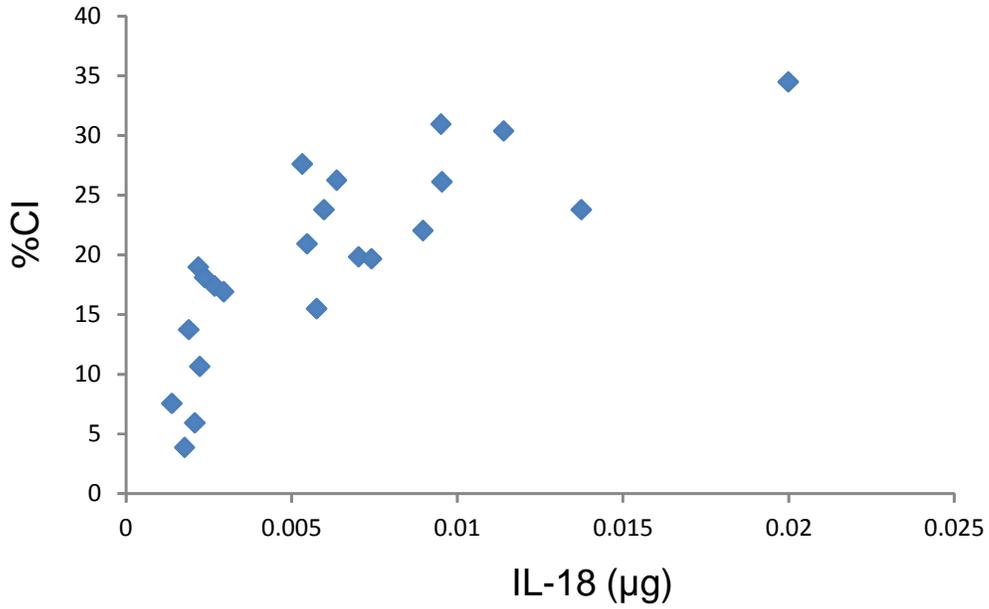

6B

| Regression/Type | Equation | $r^2$ | r | p |
|---|---|---|---|---|
| Linear | y = 1368x + 11.27 | 0.62 | 0.78 | <.01 |
| Logarithmic | y = 9.13ln(x) + 68.56 | 0.73 | 0.85 | <.01 |
| Polynomial | y = -74278x$^2$ + 2728.3x + 7.23 | 0.69 | 0.83 | <.01 |
| Power | y = 371.29x$^{0.57}$ | 0.61 | 0.78 | <.01 |
| Exponetial | y = 10.77e$^{79.21x}$ | 0.45 | 0.67 | <.01 |

**Figure 6: CI and Urine IL-18.** CI tracks 24 hr urine IL-18 across a range of values (A). A linear correlation is also observed between these 2 variables across the CI spectrum (B).

Consistent with the CI and kidney mass and CI and kidney to body mass ratio correlations, a linear fit gave excellent correlation with CI for BUN, SCr and urine IL-18. Furthermore, the CI values observed in this study span a broad range from 0.4 to 35% which more than likely falls within the dynamic range of CI observed clinically in ARPKD survivors. Finally, at best there was only ~10% difference between the r values for the linear fit vs. the best fit (see Figures 4B, 5B and 6B). A slightly "better" fit was therefore sacrificed in favor of a linear fit as long as long as the correlation met the criteria listed under Methods. As shown in Table 3, linear regressions were used to generate a family of equations for calculating CI with BUN, SCr and urine IL-18 being the variables.

| y | f(x) |
|---|---|
| CI | 0.95 * BUN (mg/dL) - 8.15 |
| CI | 53.63 * SCr(mg/dL) - 9.66 |
| CI | 1368 * urine IL-18(µg) + 11.27 |

**Table 3: CI as a Function of Biomarkers.** CI can be computed using any member of a family of equations. In these equations, the variables driving CI are BUN, SCr and 24 hr urine IL-18.

Now, if,

$$y \equiv f(x_1),$$
$$y \equiv f(x_2),$$
$$\text{and } y \equiv f(x_3),$$

then it follows from the principle of signal averaging [12] that

$$y \equiv [f(x_2) + f(x_2) + f(x_3)]/3$$

In other words, as shown in Figure 7, using more than 1 biomarker, each of which tracks CI, results in increased fidelity for estimating CI.

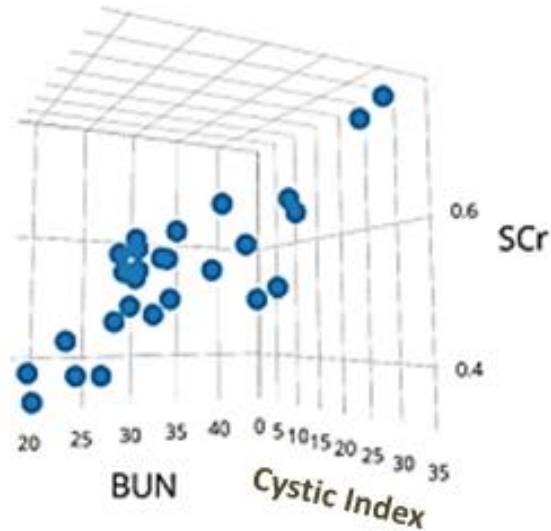

**Figure 7: CI vs a Biomarker Pair.** A 3-dimensional scattergram showing CI as a function of SCr and BUN. A robust linear correlation is observed. Including urine IL-18 in this plot would have required an additional spatial dimension.

As seen in Figure 8, we have therefore condensed the family of equations in Table 3, into a single equation/formula viz.

$$\%CI = (0.32 * BUN(mg/dL)) + (17.9 * SCr(mg/dL)) + (456 * urine\ IL18(\mu g)) - 2.18$$

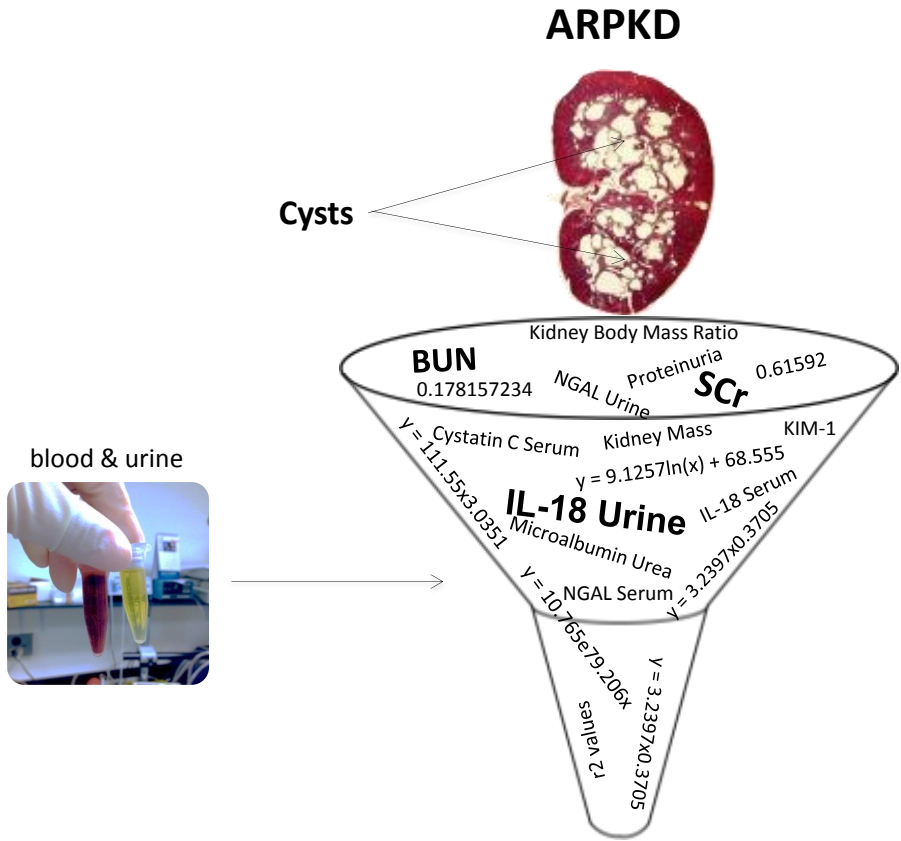

%CI=(0.32*BUN(mg/dL))+(17.9*SCr(mg/dL))+(456*urine IL18(µg)) - 2.18

*The Nieto-Narayan Formula*

**Figure 8: CI Calculator in ARPKD.** Big data–like analysis of multiple blood and urine-based biomarkers of renal injury/dysfunction yielded a calculator for estimating CI in ARPKD.

**Discussion**

Using a big data-type analysis, we herein report for the first time a quantitative relationship between CI and the levels of certain serum and urine biomarkers in a model of ARPKD; this formula can be used to estimate CI.

A genetically acquired and congenital disease, ~20-30% of ARPKD patients succumb with the first 1-2 months of life with pulmonary insufficiency secondary to renal enlargement as the primary cause of death [1-4]. For children making it past that stage, improvements in health care and disease management call for nephrectomy + dialysis or kidney transplant by ~ 10 years of age [6]. Intervention at this age is driven both by the need for reduction in flank pain due to highly enlarged kidneys as well as severe renal insufficiency. Formation and expansion of fluid-filled cysts drives both renal enlargement and renal insufficiency [5]. A hallmark feature of ARPKD is that cyst formation and renal enlargement precede the decrease in renal function [6]. This means that when a precipitous decline in renal function is observed, cystic damage to the renal parenchyma is severe.

Today, ARPKD patients are subjected to renal imaging every so often as a means of tracking cyst formation and expansion. Although ultrasound is currently the imaging modality of choice for a diagnosis of ADPKD, its utilization for assessing disease progression is limited; i.e. ultrasound is a diagnostic but not prognosticative tool. Ultrasonography is highly operator-dependent and produces images that are less sensitive and reproducible than computed tomography (CT) or magnetic resonance imaging (MRI) [8]. Furthermore, ultrasound cannot reliably detect small but significant changes in kidney size or small cysts that are less than one centimeter in diameter. Ultrasound based measurements of kidney volume are calculated by utilizing parameters based on a modified ellipsoid formula. Problems related to the use of the ellipsoid formula include inter-operator variability when determining the maximal anterior-posterior diameter, the variability induced by patient respiration, and the lack of uniform cyst expansion throughout the kidney in this disease, making the ellipsoid a less than perfect estimation of size [8]. The contributions of these variables typically result in an overall underestimation of kidney volume. Measuring cyst volume is more difficult compared to total kidney volume. Due to the irregular shape and enormous number of cysts, only CT or MRI methods using the integral or voxel counting method can accurately measure cyst content [8]. Unfortunately such contrast-enhanced or even simple scanning modalities are expensive, time-consuming and somewhat impractical in the pediatric population. Certainly, ARPKD children and their parents do not savor the experience of undergoing repeat contrast MRI or CT.

We have previously measured the levels of several serum and urine-based biomarkers a model in a well-accepted mammalian model of ARPKD [9,10]. The PCK rat has a mutation in the *PKHD1* gene and exhibits all the hallmark features of human ARPKD and congenital liver fibrosis [9] i.e fibrocystic human disease. In the present study, we adopted a big data-like approach to identify and quantify the relation between these biomarkers and CI. Our findings suggest that of the biomarkers studied, BUN, SCr and urine IL-18 are of particularly useful in that they correlate linearly with CI across a broad range of cystic pathology. Implicit in this observation is that these biomarkers can potentially be used to estimate CI from the very early through the late disease stage. Furthermore, since a signal averaging protocol was used to combine these 3 biomarkers, each of which tracks CI, the resulting formula is expected to provide increased fidelity for estimating CI. Importantly, from a pediatric patient perspective, serum and urine (which are routinely collected in this population)-based tests bring enhanced compliance when compared to time-, labor-, expense-intensive and uncomfortable longitudinal imaging protocols. Finally, use of a 24 hr urine sample, while somewhat cumbersome, eliminates any confounding factors with spot urine collection.

There are some limitations to our findings. The present study used a database stemming from urine, serum and renal samples from 13.5 week old PCK rats. Our empirical formula can benefit from further refinement by use of larger database that incorporates samples from several different timepoints in the PCK rat model as well as other mammalian models of ARPKD such as the B6.129S6-*Pkhd1$^{tm1Cjwa}$*/J (Jackson Labs) mouse model. More importantly, the formula will eventually need to be validated in ARPKD patients. Serum and urine samples from nephrectomy candidates will need to be obtained and entered into the calculator and the calculated CI correlated with the measured CI from the discarded kidney.

These limitations notwithstanding, our results form the foundation for developing a calculator in ARPKD along the lines of other existing calculators for renal and liver diseases such as the modified diet in renal disease (MDRD), chronic kidney disease-epidemiology collaboration (CKD-EPI), FIB-4 and aspartate aminotransferase to platelet ratio index (APRI) calculators [13-15]. Such an empirical ARPKD calculator will not only represent a patient-friendly and relatively inexpensive method to track disease progress and aid in the management of this population but can also be used in clinical trials of drugs that work by reducing the formation and growth of cysts and therefore eventually retard renal dysfunction.

**Acknowledgements**

The contributions of Brian Huang, Ping Zhou, Latha Paka and Prani Paka in obtaining the source data are gratefully acknowledged.